\documentclass{aa}

\usepackage[varg]{txfonts}
\usepackage{graphicx}
\usepackage{natbib}
\bibpunct{(}{)}{;}{a}{}{,} 
\usepackage{amssymb}
\usepackage{amsmath}
\usepackage{multirow}
\usepackage[colorlinks=true,     linkcolor=blue, citecolor=blue, filecolor=blue, urlcolor=blue]{hyperref}

\begin{document}

\title{Implementation of thermal conduction energy transfer models in the Bifrost solar atmosphere MHD code}

\author{George Cherry\thanks{\email{georgche@uio.no}} \and Mikolaj Szydlarski  \and Boris Gudiksen}

\institute{Rosseland Centre for Solar physics,Universitetet i Oslo, Sem Sælands vei 13, 0371,Oslo, Norway \and Institutt for Teoretisk Astrofysikk, Universitetet i Oslo, Sem Sælands vei 13,0371,Oslo, Norway}

\abstract{Thermal conductivity provides important contributions to the energy evolution of the upper solar atmosphere, behaving as a non-linear concentration-dependent diffusion equation. Recently, different methods have been offered as best-fit solutions to these
problems in specific situations, but their effectiveness and limitations are rarely discussed.}{ We have rigorously tested the different implementations of solving the conductivity flux, in the massively parallel magnetohydrodynamics code, Bifrost, with the aim of specifying the best scenarios for the use of each method.}{We compared the differences and limitations of explicit versus implicit methods, and analyse  the convergence of a hyperbolic approximation. Among the tests, we used a newly derived first-order self-similar approximation to compare the efficacy of each method analytically in a 1D pure-thermal test scenario.}{We find that although the hyperbolic approximation proves the most accurate and the fastest to compute in long-running simulations, there is no optimal method to calculate the mid-term conductivity with both accuracy and efficiency. We also find that the solution of this approximation is sensitive to the initial conditions, and can lead to faster convergence if used correctly. Hyper-diffusivity is particularly useful in aiding the methods to perform optimally.}{ We discuss recommendations for the use of each method within more complex simulations, whilst acknowledging the areas where there are opportunities for new methods to be developed.}


\maketitle


\section{Introduction} \label{sec:intro}
Our closest star continues to offer a wealth of enigmatic and mystifying phenomena that forge an exciting laboratory for physics relating to plasma, particles, nuclear, fluid dynamics, and thermodynamics, to name a few. Since the early 1950s, researchers have persevered to connect theory to observations, and vice versa. The development of faster and larger supercomputers, as well as better observational equipment over the past decade, has further accelerated this progress, and yet some of the oldest questions, such as the coronal heating problem, still remain unanswered. If such questions are to be answered, the need to capture the complexity of varying physical processes that occur across the full extent of the solar atmosphere must be met. \\

To that end, \verb|Bifrost| is a 3D massively parallel magnetohydrodynamics (MHD) code, aiming to simulate stellar atmospheres from the upper convection zone through to the lower corona in detail. First and foremost, \verb|Bifrost|'s aim is realism, and as such the precise implementation of additional physics such as radiative transfer, the generalised Ohm's law, and thermal conduction is a key attribute in the development of the code. We focus on thermal conduction, which causes additional transport of heat in the upper solar atmosphere (chromosphere and corona). The contribution manifests itself in the internal energy equation,
\begin{eqnarray}
    \frac{\partial e}{\partial t} = - \nabla \cdot e\mathbf{u} - p \nabla \cdot \mathbf{u} + Q, \label{eqn: energy}
\end{eqnarray}
where $Q$ represents the divergence of the heat flux from thermal conduction.

\citet{Spitzer1962} showed that the heat flux behaves as a non-linear diffusion equation in the direction of the magnetic field, such that
\begin{eqnarray}
    Q &=& -\nabla_\parallel \cdot (\mathbf{q} ),   \\
    \mathbf{q} &=& -\kappa_0 T^{5/2}\nabla_{\parallel} T,
    \label{Eqn: spitzer}
\end{eqnarray}
where $\kappa_0 \sim 10^{-6} \text{ erg cm}^{-1} \text{ s}^{-1} \text{ K}^{-7/2} $ is the Spitzer conduction coefficient. A full decomposition of the conductive flux, including perpendicular and transverse components, can be found in \citet{Braginskii1965}. In the corona, where the temperatures are largest, and hence the conductive flux has the most effect on the internal energy, the perpendicular and transverse terms are found to be much smaller than the parallel terms. In the chromosphere, however, the perpendicular contribution becomes comparable to the parallel conductive flux \citep{Navarro2022}. Nevertheless, perpendicular and transverse values are on the same scale as numerical diffusion within \verb|Bifrost| and so these components are not currently implemented \citep{Gudiksen2011}. \\

When discretised explicitly using a numerical scheme, for example as
\begin{eqnarray}
    \frac{\partial e[i,n]}{\partial t} &= &\frac{e[i,n]-e[i,n-1]}{\Delta t}, \\
    \nabla q[i,n] &=& \frac{q[i,n] - q[i-1,n]}{ \Delta x},
\end{eqnarray}
in the 1D case, the second-order diffusive operator limits the Courant condition to scale with $\Delta x^2$. This greatly limits the suitable time-stepping ($\Delta t$) for finer grids, which, even using parallel codes with high-performance computing, requires large amounts of memory and CPU hours to compute. Simulating the long-term evolution of large-scale events, with timescales on the order of solar hours, or even days, becomes time-consuming and error-ridden. Instability within explicit methods is also difficult to manage and requires large artificial diffusive parameters to maintain it. \\

\verb|Bifrost| employs two other methods of solving thermal conduction. The first uses the Crank-Nicholson implicit method described in \citet{Gudiksen2011},
\begin{equation}
    \frac{e[n+1]-e^*}{\Delta t} = \theta \nabla_\parallel \cdot \mathbf{ q}[n+1] + (1-\theta) \nabla_{\parallel} \cdot \mathbf{q}[n].
    \label{Eqn: Implicit}
\end{equation}
The quantities at time-step $n$ are calculated before the MHD time-step, starred quantities are produced after the MHD time-step, and $[n+1]$ quantities are deduced implicitly, using a multi-grid method \citep{MALAGOLI}. Finally, the energy term is overwritten altogether. This method is known to be unconditionally stable, but still subject to other numerical artefacts and error build-up. Nonetheless, it is a more robust method than the explicit version, offering larger time-steps with stability, but at the cost of more computing time due to the complex machinery of the method.  \\

The second method implements the hyperbolic PDE approximation proposed by \citet{Rempel2017},
\begin{eqnarray}
    \frac{\partial \mathbf{q}}{\partial t}&=& \frac{1}{\tau} \left(- \mathbf{q}  -\kappa_0 T^{5/2} \nabla_{\parallel} T \right), \label{Eqn: wave}\\ 
    \frac{\partial e}{\partial t} &=& \text{...} - \nabla \cdot \mathbf{q}, 
\end{eqnarray}
which converges on the usual parabolic Spitzer equation  (\ref{Eqn: spitzer}) over time. For $\tau>0$, this is the equivalent of solving a wave equation for $T$, and as such the solution is a wave, with a propagation speed of $c = \sqrt{\kappa/\tau}$, when $\kappa$ and $\tau$ are constant. The wave method, as it will now be referred to, is robust, reducing any instabilities that occur, and faster to compute than the above implicit method, as it is still calculated explicitly. It must be noted that this method only converges on the correct flux values in time, relative to the convergence timescale $\tau$. Care must be taken with the choice of $\tau$. Since the Courant condition now also depends on $1/\tau$, the convergence rate must not become too small or the system will become unstable. \\

It can be seen that each method has its positives and negatives. The following sections discuss the utilisation of these methods in the context of MHD codes.  Section \ref{sec: Implem} discusses the structure of the numerical schemes and specifics of the machinery in the \verb|Spitzer| module within \verb|Bifrost|. Sections \ref{sec: Methedology} and \ref{sec:resul} rigorously test the capabilities of each method, through two idealised tests and comparing them to analytical solutions. We conclude with a discussion on how to use these methods optimally when applied to more complex MHD simulations and remark on the development of this module for the future.

\section{Numerical schemes} \label{sec: Implem}
Since thermal conduction is not a driver of the evolution of any MHD quantities except internal energy, its contribution may be calculated independently of the main MHD calculations in a simulation. For the explicit and wave methods, the calculated conduction component is simply added to the current $\frac{\partial e}{\partial t}$ before the next time-step. For the implicit method, the final internal energy, $e$, is overwritten to include the effect of the conductive flux. Thus, whilst \verb|Bifrost|'s main calculation works on an explicit Hyman or Runge-Kutta method, the thermal conduction calculations are free to be calculated in whichever method suits the simulation. It is important to understand how the methods' different behaviours affect the outcome of the \verb|Spitzer| module, and ultimately the heat transfer in any simulation.

\subsection {Explicit method}
The explicit method is a first-order Euler method. The greatest cause of instability for this method is the Courant condition, $C \sim \frac{\Delta t}{\Delta x^2}$. We can split this into three parts for each derivative calculated; two spatial ($\nabla T$, $\nabla \cdot \mathbf{q}$) and one temporal ($\frac{\partial e}{\partial t}$). Firstly, in order to ensure stable handling of high temperatures within the code (important for the corona, where $T$ can reach above $10 \text{ MK}$), we calculate the temperature gradient of $\log(T)$, such that the flux term becomes
\begin{equation}
\mathbf{q} =  \kappa(T) \nabla_\parallel \log T = \kappa_0 T^{7/2} \mathbf{\hat{b}} (\mathbf{\hat{b}} \cdot \nabla) \log T,
\end{equation}
where $\mathbf{\hat{b}} = \mathbf{B}/|\mathbf{B}|$ is the unit vector for the magnetic field. This also helps to lessen the instability within the first spatial derivative due to steep gradients. The second derivative essentially concerns  the rate of change of the temperature gradient, $\nabla^2 T$. This can cause issues around, for example, shock regions, as is seen in Sect. \ref{sec:resul}. Finally, the time derivative must heed that the total change in energy per time-step does not exceed the amount of energy in the system, or else the numerical model will break and the solution will become unphysical. \verb|Bifrost| monitors this relation through the condition
\begin{equation}
\kappa(T) = \kappa_0 T^{7/2}  < \beta \left(\frac{\Delta x^2}{\Delta t}\right) e,  \label{eqn: expl_lim}
\end{equation}
for $\beta \in [0,1)$. This is observed before the final spatial derivative is taken such that problems due to large gradients in T can also be monitored this way. In \verb|Bifrost|, $\beta = 0.7$, which allows for energy loss from other processes.

\subsection{Implicit method}
This method is an adaptation and extension to 3D of the non-linear multi-grid scheme presented in \citet{MALAGOLI}. Firstly, the hydrodynamic energy equation is calculated, alongside an explicit approximation, very similar to above, for $\kappa(T)\nabla T$ at the current time-step, which is treated as a source term for the implicit calculation. The weight of this first step to the total conductive flux contribution is decided by $\theta \in [0,1]$. For $\theta = 0$, we recover the explicit method. As such, it is important that $\theta$ is suitably large that instabilities are completely smoothed for larger time-steps. If $\theta \sim 1$ however, the implicit method will diffuse the solution too much unnecessarily. Here, we used $\theta=0.9$, although the results for the following tests do not change significantly for $\theta >0.5$, where the instability of the explicit part begins to dominate for $\theta < 0.5$. \\

The Crank-Nicholson implicit calculation uses a non-linear multi-grid solver for the inversion of the operators, which accelerates the convergence of the relaxation iterations. It performs corrections to the calculated solution on a selection of fine to coarser grid patterns, in order to reduce errors with wavelengths comparative to each grid size \citep{Millar2003}. Once the coarsest resolution has been reached, the resolution is once again increased back to the original resolution, in a pattern known as a V-cycle, due to its implied shape. For each grid resolution, $h$, the interaction between it and the next coarsest level, $H = 2h$, uses the full approximation scheme (see \citet{Brandt2011} for a detailed explanation):
\begin{enumerate}
    \item At level $h$, the solution to the implicit non-linear problem  $Lu=f$, where $u$ is unknown, is denoted as $L^hu^h=f^h$.
    \item After a few Gauss-Seidel relaxation sweeps, the smoothed solution is approximated as $\tilde u^h$.
    \item A correction term, $v^h$, is needed, such that $v^h+\tilde u^h =  u^h$.
    \item The residual, $r^h = L^h(\tilde u^h + v^h) -L^h\tilde u^h$, is injected into the coarser grid level, $H$.
    \item Using an interpolation scheme, $I_h^H$, the following values can be interpolated to the new resolution:
        \begin{itemize}
            \item $u \to u^H = I_h^H \tilde u^h + v^h$,
            \item $f \to f^H = L^H(I_h^H \tilde u^H ) + \hat I_h^H \tilde u^h$, where $I$ and $\hat I$ pertain to separate interpolation processes,
        \end{itemize}
        such that the linear problem becomes $L^Hu^H=f^H$.
    \item An approximation, $\tilde u ^H$, is calculated for the above using Gauss-Seidel sweeps. The new correction term can be approximated by $v^H = \tilde u^H - I_h^H \tilde u^h$.
    \item The coarse correction term is injected back into the finer grid scheme through interpolation, and used to update the finer solution approximation,
    $\tilde u^h_{\text{NEW}} = \tilde u^h + \hat I_H^hv^H$.
    
\end{enumerate}
Between steps 6 and 7, it is possible to include more levels of coarser grids, by injecting the new residual to lower grids and repeating steps 5 and 6, before updating each solution approximation in the finer grids. By the coarsest grid, the calculations must be processed on only one core in order to keep communication costs low in relation to the computations. Each level is iterated for a select number of repetitions and the full cycle is also repeated as desired. The more repetitions, the faster the solution will converge, and so it is beneficial to iterate each level multiple times. The coarser grids may be repeated many times without a large computational cost, and as such tend to be repeated more than the finer grids. \verb|Bifrost| has four levels, with each level half the resolution of the level above it. For the purpose of this paper, the iterations of each level are 4,4,4 and 16, respectively, with the whole cycle being repeated twice. \\

\subsection{Wave method}
By rearranging equation \ref{Eqn: wave}, we recover the hyperbolic equation for calculating $q$, given by $\mathbf{q} = q\mathbf{\hat{b}}$, proposed by \citet{Rempel2017},

\begin{equation}
    \frac{\partial q}{\partial t} = \frac{1}{\tau} \left( - \frac{q^*}{f_{\text{sat}}} - q \right),
    \label{Eqn: wave_1}
\end{equation}
and where $q^* =\kappa_0 T^{5/2} (\mathbf{\hat b} \cdot \nabla) T$ recovers the original Spitzer equation. Rempel discusses appropriate values for the saturation rate ,$f_{\text{sat}}$, and convergence rate, $\tau$, as
\begin{eqnarray}
    f_{\text{sat}} &=& 1 + \frac{|q*|}{1.5 \rho C_S^{3}}, \\
    \tau &=& \left(f_{\text{CFL}} \frac{\Delta x_{\min}}{\Delta t}- | v | \right)^{-2}\frac{f_{\text{sat}}\kappa_0 T^{7/2}}{e}, \label{Eqn: tau}
\end{eqnarray}
where $C_S=\sqrt{\gamma p / \rho}$ is the ideal gas sound speed. It is possible to omit the saturation factor, $f_{\text{sat}}$, in equation \ref{Eqn: wave_1} for faster convergence, but this also decreases the stability of the method \citep{Navarro2022}, as is discussed in Sect. \ref{sec:resul}. The choice of the equation for $\tau$ is flexible and somewhat depends on the code. For example, \citet{Navarro2022} instead opt for a constant $\tau = 4 \Delta t$ throughout, whereas \citet{warnecke2020} relate $\tau$ to the Alfvén velocity and the conductivity. In the latter case, lower and upper limits are both imposed on $\tau$ for similar reasons to the ones discussed below. $f_\text{CFL}$ was taken to be 0.7 for our tests. However, the choice, $f_{\text{CFL}} \in [0,1]$, has a negligible effect in these tests, since we have ignored dynamics (e.g spatially varying velocities) and because the dominating term is proportional $(\Delta t )^{-2}$.\\

This method creates an additional calculation that must be taken at each time step, whereby the new flux term, $q$, is calculated from the time derivative in equation \ref{Eqn: wave_1}. Therefore, for time step $[n]$, there are two operations:
\begin{enumerate}
    \item Add the flux calculated at the previous time to the energy equation,
    \begin{equation}  e[n+1] = e[n] + \Delta t ( ... - \nabla \cdot (\mathbf{q}[n])). \end{equation}

    \item Calculate the flux to be used at the new time using equation \ref{Eqn: wave_1},
    \begin{equation} 
    q[n+1] = q[n] + \Delta t \left[ \frac{1}{\tau} \left(-\frac{q^*[n]}{f_\text{sat}} - q[n]\right)\right]. \label{Eqn: new_q}
    \end{equation}
\end{enumerate}
Therefore, the stability of a fourth derivative calculation, in addition to those already discussed in the explicit method,  must be considered. From equation \ref{Eqn: tau}, it is clear that $\tau$ scales with $(\Delta t)^2$, and as such, the Courant condition for equation \ref{Eqn: new_q} scales with $1/\Delta t$. In other words, the system becomes unstable for \textbf{smaller} time-steps. Therefore, it is important that the wave implementation contains a 'throttle' in the form of a lower limit on $\tau$, 
\begin{equation}  
\tau_{\min} = \alpha \Delta t, \label{eqn: throttle}
\end{equation}
where the choice of $\alpha$ is discussed more in Sect. \ref{sec:resul}.\\

\subsection{Adaptive time-stepping}

By default, Bifrost varies the time-step of the current running simulation,  by monitoring the overall Courant condition. When additional physics modules, such as thermal conduction, are involved, it is important that the local Courant condition, and any other conditions for the chosen method, are also considered.  The limit in equation \ref{eqn: expl_lim} for the explicit method is an example of this. We can compare the tests described below for both constant and variable time-stepping, in order to understand where such limiting factors may come in to play within a simulation.

\begin{figure*}
    \centering
    \includegraphics[scale=1.2]{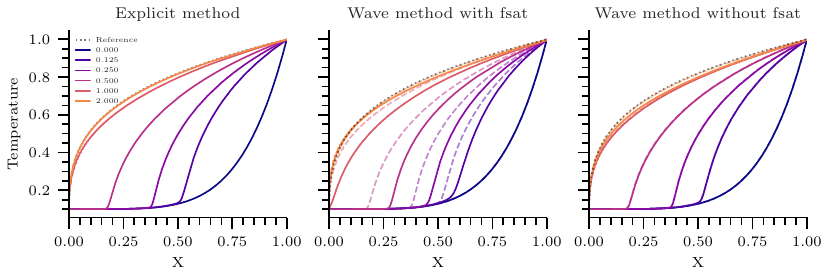}
    \caption{Diffusion of the wave method with and without saturation for $\Delta t = 10^{-6}$. The explicit solution for $\Delta t = 10^{-7}$ is indicated by dashed lines in the final two panels. The solid lines indicate evolution at times $t=0.0, \, 0.125, \, 0.25, \, 0.5, \, 1.0, \text{ and } 2.0.$ The dotted line references the analytical solution for convergence.}
    \label{fig:1 D_comp}
\end{figure*}

\section{Methodology}
\label{sec: Methedology}

\begin{figure}
\centering
    \resizebox{\hsize}{!}{\includegraphics{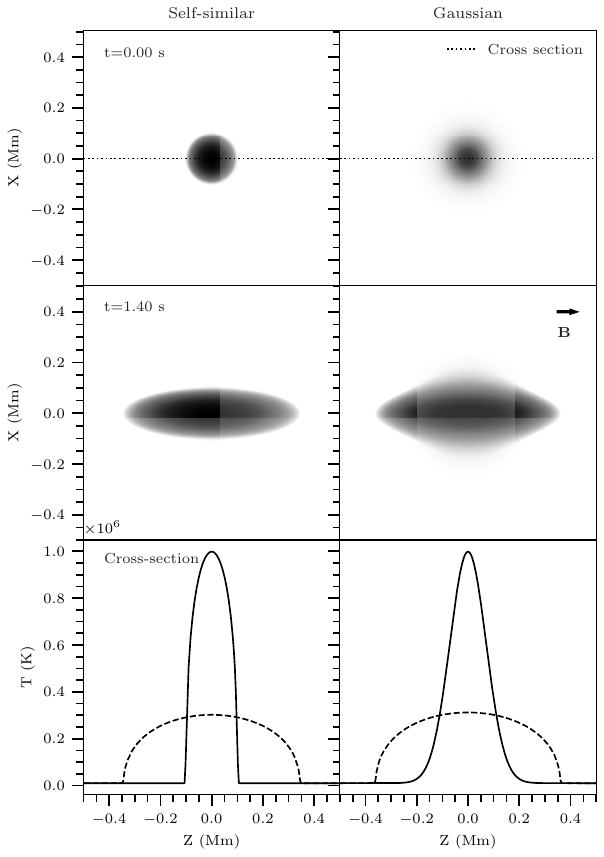}}
    \caption{Evolution of the self-similar diffusion test for Gaussian and self-similar initial conditions. At $t=1.40 \text{ s}$, the profile lies in the $\chi~t~\gg~1$ regime and both initial profiles show a self-similar diffused profile, albeit for different $\phi_0$. The solutions are run with the explicit method for $\Delta t= 10^{-5} \text{ s}$. We note that different scales are used for the colour maps between $t=0 \text{ s and } t=1.4 \text{ s}$.}
    \label{fig: ss_ic}
\end{figure}

We now present two tests with which to validate and analyse the methods. The first is a 1D convergence test, described before by \citet{Rempel2017} and also used by \citet{Navarro2022} in their testing of the Mancha3D code. Thus, this test is suitable for validating the implementation and efficacy of the wave and explicit methods. We note that the implicit multi-grid solver is tuned for at least a 2D set-up, so was not included for the 1D test. The second test calculates the diffusion of a Gaussian temperature profile in a 2D set-up. The numerical results are then compared to an analytical first-order approximation by \citet{Furuseth2024}. Both tests require a purely thermal setup, such that all MHD time derivatives, except energy, remain zero. Examples of these tests can be seen in Figs. \ref{fig:1 D_comp}and \ref{fig: ss_ic}, respectively.

\subsection{1D convergence test}

We considered the 1D non-linear heat diffusion equation,
\begin{equation}
    \frac{\partial T}{\partial t} = \kappa_0 \frac{\partial }{\partial x} \left ( T^{5/2} \frac{\partial T}{\partial x} \right),
\end{equation}
with an initial temperature profile of
\begin{equation}
    T_0(x)=0.1 + 0.9x^5,
\end{equation}
and fixed boundaries such that $T(0)=0.1$ and $T(1)=1$ for all $t$. For $t>>0$, the temperature solution converges ($\frac{\partial T}{\partial t} = 0$) to the asymptotic stationary solution 
\begin{equation}
    T=\left[ 0.1^{7/2} + (1-0.1^{7/2})x \right]^{2/7}, \label{eqn: asympt_sol}
\end{equation}
where $\kappa_0$ controls the speed of convergence. \\

Our domain contained $256$ grid points, extending across $x~=~[0,1]$. We normalised the set-up to unity, providing a constant 1D magnetic field, $B_x =1.0$, and density, $\rho=1.0$. Internal energy and pressure were calculated in the ideal gas regime, such that they became directly proportional to $T$. We note that the above solution pertains to a time evolution in $T$,
\begin{equation}
    \frac{\partial T}{\partial t} = - \nabla_\parallel \cdot \mathbf{q}(T),
\end{equation}
whereas our implementation was directly solved in the energy equation, and so pertained to a time evolution in $e$ (see equation \ref{eqn: energy}),
\begin{equation}
    \frac{\partial e}{\partial t} = \frac{\rho}{\gamma-1}\frac{\partial T}{\partial t} = -\nabla_\parallel \cdot \mathbf{q}(T),
\end{equation}
in the ideal gas and normalised regime.
Therefore, given a constant density, the speed, $\kappa_e$, at which the test within \verb|Bifrost| converged to the solution (\ref{eqn: asympt_sol}) is related to the Spitzer coefficient $\kappa_0$ by
\begin{equation}
    \kappa_e = \frac{\rho}{\gamma -1} \kappa_0,
\end{equation} 
where we have normalised the Boltzmann constant, $k_B=$, ion mass, $m_H$, and mean molecular weight, $\mu$, to 1.0. Furthermore, we set $\gamma = 5/3$ and $\kappa_0 = 1.0$ to agree with the tests in \citet{Rempel2017} and \citet{Navarro2022}.\\

\subsection{Self-similar diffusion test}

The self-similar solution to a non-linear diffusion equation with concentration-dependent coefficients in up to three dimensions, has been  studied analytically by \citet{Pattle1959}. Analytical descriptions of the evolution and convergence from an instantaneous point source with a zero background concentration, and finite boundary conditions ($T(r) = 0$ for finite $r$) are provided. In the case of thermal conduction, this concentration is taken to be temperature, $T$. Taking the dimension as one, since  diffusion is only along the magnetic field, this leads to the following behaviour:

\begin{itemize}
\item Evolution of a point source with initial central value $T_0$ and width $R_0$ at the foot-points

\begin{align}
    T(r,t) &= 
    \begin{cases}
        T_0\left(1+\chi t\right)^{-2/9}
        \left(1-\dfrac{r^2}{R(t)^2}\right)^{2/5} 
        ,  & \text{if $r<R(t)$} \\ 
        0~, & \text{otherwise}.
    \end{cases} \label{eq_T}
    \\
    R(t) &= R_0\left(1+\chi t\right)^{2/9}, \label{eq_R} \\
    \chi &= \frac{9}{5} \frac{2\kappa_0T_0^{5/2}}{R_0^2},  \label{eq_chi}
\end{align}
 where $T_0$ and $R_0$ are determined by the total surface integral $\phi_0 = \frac{T_0 R_0}{B}$ for the Beta function $B=B(\frac{1}{2},\frac{7}{2})$. \\

\item Convergence when $\chi t >> 1$ of \textbf{any distribution} with total surface integral $\phi_0$ \\
\begin{align}
    T(r<R(t),t) &\to 
        T_0\left(\frac{5}{9}\frac{(B \phi_0)^2}{2\kappa_0t}\right)^{2/9}
        \left(1-\dfrac{r^2}{R(t)^2}\right)^{2/5}  \label{eq_T_conv} \\
    R(t) &\to \left(\frac{9}{5}\kappa_0 (B \phi_0 )^{5/2} t\right)^{2/9}.\label{eq_R_conv}
\end{align}
\end{itemize}

Equations \ref{eq_T_conv} and \ref{eq_R_conv} provide an important result of the solution, that the converged self-similar solution does not depend on the initial central value of the distribution, and as such, any distribution that has the same $\phi_0$ as the given instantaneous point source, and becomes zero in a finite domain, will in fact converge on the same shape. \\

This solution has been used already in \verb|Bifrost| to test and validate the implementation of ambipolar diffusion \citep{Fernando2022}. However, since the physical quantity we considered is temperature, the system became unphysical at $0 \text{ Kelvin}$. We therefore considered a small background temperature, $T_\infty$, such that $T_\text{tot} (r,0) = T_\infty + T (r,0)$, and compared to a first-order modification of the self-similar solution in the regime $T_\infty << T_0$ \citep{Furuseth2024}:
\begin{align}
    \max[T_\text{Tot}^{(1)}(t)] &= 
    \begin{cases}
        T_\infty + T_0\left[1+\chi t\left( 1 + \frac{T_\infty}{T_0}\right)^{5/2}\right]^{-2/9}
          &, \text{if $\chi t \ll 1$} \\ 
        T_\infty + T_0\left[1+\chi t\left(\tfrac{\phi_{\text{Tot}}}{\phi_0}\right)^{5/2}\right]^{-2/9} 
          &, \text{if $\chi t \gtrsim 1$} \\
        \to  T_\infty + \left(\tfrac{5}{9}\tfrac{(B \phi_\text{Tot} ) ^2}{2\kappa_0 t} \right)^{2/9}
        &, \text{if $\chi t \gg t$}.
    \end{cases} \label{eq_T_1}
    \\
    R^{(1)}(t) &= 
    \begin{cases}
        R_0 \left [ 1+ \chi t \left( 1 + \tfrac{T_\infty}{T_0} \right)^{5/2} \right ] ^{-2/9}
        &, \text{if $\chi t \ll 1$} \\
        R_0 \left [ 1+ \chi t \left(\tfrac{\phi_{\text{Tot}}}{\phi_0} \right)^{5/2} \right]^{2/9}
        &, \text{if $\chi t \gtrsim 1$} \\
        \to \left( \tfrac{9}{5} 2 \kappa_0 (B \phi_\text{Tot})^{5/2} t \right)^{2/9}
        &, \text{if $\chi t \gg 1$}.
    \end{cases} \label{eq_R_1}
\end{align}

Our set-up took a Gaussian temperature profile in a 2D domain, with an initial background temperature of 1\% of the peak total temperature $T_{\text{Tot}}$,
\begin{equation}
    T=9.9\times 10^5 \exp\left(-\frac{x^2+z^2}{0.01 ^ 2}\right) + 1 \times 10^4.
\end{equation}

It should be noted that the analytical solutions pertain to an initial condition with a self-similar shape (see equation \ref{eq_T}), but we show (see Fig. \ref{fig: ss_ic}) that in the regime important to us, $\chi t \gg 1$, the Gaussian initial condition converges to the self-similar solution with the same perturbed area, $\phi_0$, as with the converged zeroth-order equations, \ref{eq_T_conv} and \ref{eq_R_conv}. In our case, that self-similar profile uses $R_0 = 0.1 \text{ Mm}$, $T_0 = 9.9 \times 10^{5} \text{ K}$, and $T_{\infty} = 1 \times 10^4 \text{ K}$ \\

In our ($1 \times 1$) Mm domain, centred about (0,0), and split into ($256 \times 256$) grid points, we imitated coronal conditions \citep{Carlsson2019} with low density, $\rho = 10^{-15} \text{ g} \ \text{cm}^{-2}$, following the ideal gas regime, and weak magnetic field in one direction, $B_z =0.316 \text{ G, } B_{x,y}=0 text{ G}$. Whereas the magnitude of the magnetic field is arbitrary, the density value affects the speed of diffusion, as discussed in the previous test, but this time without unity normalisation. Therefore, when comparing to the analytical solution, we modified the analytical $\kappa_0 \to \kappa_\alpha$,
\begin{equation}
    \kappa_a  = \frac{(\gamma - 1) \mu m_h}{k_B \rho} \kappa_0,
\end{equation}
where $k_B = 1.38 \times 10 ^{-16} \text{ erg K}^{-1}$ is the Boltzmann constant, $m_h = 1.67 \times 10^{-24} \text{ g} $ is the ion mass, and $\mu = 0.5$ relates to full hydrogen ionisation, along with $\gamma=5/3$. The plasma was initially at rest ($\mathbf{u}=0$), and remained at rest throughout, relating to pure thermal diffusion. The boundary conditions for the domain were periodic. Where there was a need to analyse diffusion over a longer time period, the domain was extended to ($2 \times 2$) Mm with ($512 \times 512$) grid points, such that the resolution remained the same and the boundaries never directly interacted with the profile.\\

\section{Results}
\label{sec:resul}

\begin{figure*}
    \centering
    \includegraphics[scale=1.2]{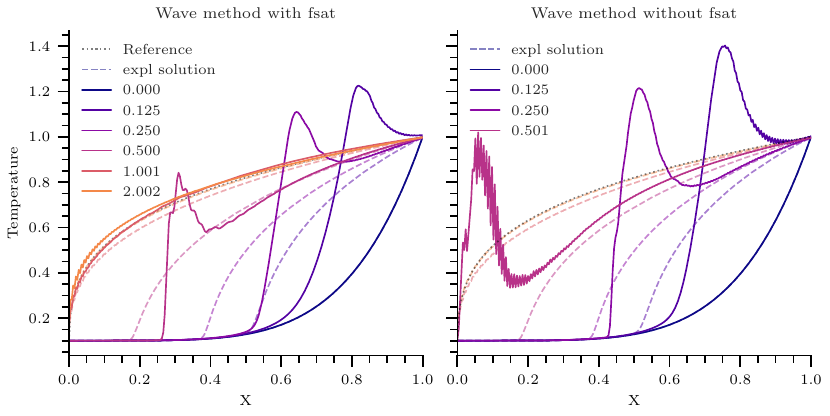}
    \caption{Diffusion of the wave method with and without saturation for non-constant $\Delta t$. The explicit solution is indicated by dashed lines.}
    \label{fig: 1D_vardt_comp}
\end{figure*}

\subsection{Accuracy}
\begin{figure}[h!]
    \centering
    \resizebox{\hsize}{!}{\includegraphics{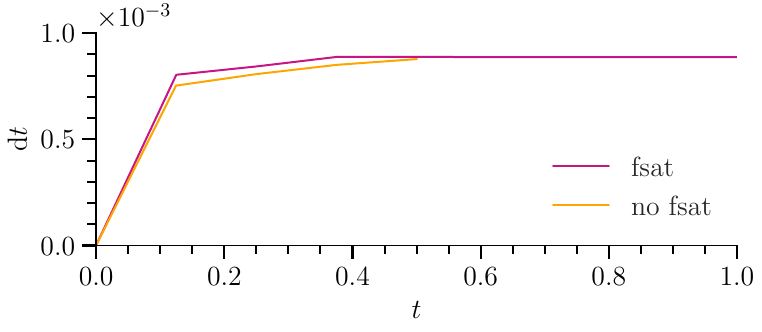}}
    \caption{ Evolution of $\Delta t$ in time for the adaptive time-stepping wave method in the 1D test, starting at $\Delta t=$1e-6.}
    \label{fig: 1D_vardt}
\end{figure}

\begin{figure}[h]
    \centering
    \resizebox{\hsize}{!}{\includegraphics[scale=1.2]{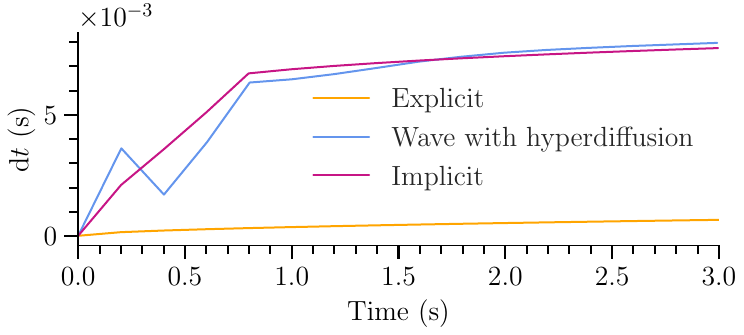}}
    \caption{$\Delta t$ in the adaptive time-stepping version of each method, with a starting $\Delta t$ of $10^{-5} \text{ seconds}$.}
    \label{fig: vardt_comp}
\end{figure}

Figures \ref{fig:1 D_comp} and \ref{fig: ss_ic} give examples of the two tests described above. \verb|Bifrost|'s explicit solution to the 1D test, agrees completely with the results of \citet{Rempel2017}, for the case $\tau=0$, which corresponds to the normal parabolic equation. We validate that the wave method without saturation factor $f_{\text{sat}}$ (right panel)  converges quickly to the parabolic solution, some time before $t=0.125$, since the results are comparable to those of the explicit method. In contrast,  with added saturation (middle panel), the wave method lags behind the explicit solution throughout, but still achieves convergence by $t = 2$. The benefits of the saturated method become apparent in Figs. \ref{fig: 1D_vardt_comp} and \ref{fig: 1D_vardt} for variable time-stepping. Where the unsaturated version (right panel) breaks down by $t= 0.5$ due to large instabilities, the saturated version is able to recover and maintain stability\footnote{The final iterations will always have some ringing due to the forced fixed boundary conditions, an unnatural setup for Bifrost} by the convergence time of $t=2.0$. We can see in Fig. \ref{fig: 1D_vardt} that the $\Delta t$ values in both versions steadily rise above $8 \times 10^{-4}$. The larger instability within the unsaturated method causes the time-step to rise slower than for the saturated version, which reaches the maximum stable $\Delta t \, (\approx 9 \times 10^{-4})$ by $t= 0.5$. The unsaturated method, can no longer control the instability however, and the simulation crashes just after $t=0.5$. This method would require a stricter control condition, lowering the maximum possible $\Delta t$ for stable runs. For this reason, \verb|Bifrost| defaults to using the saturated version, since the main benefit of the wave method is larger time-stepping over the explicit method, and a faster calculation time than the implicit method, which, as is described later (see Fig. \ref{fig: vardt_comp}), has a similar maximum time-stepping available as the saturated wave method. \\

\begin{figure*}
    \centering
    \sidecaption
    \includegraphics[width=12cm]{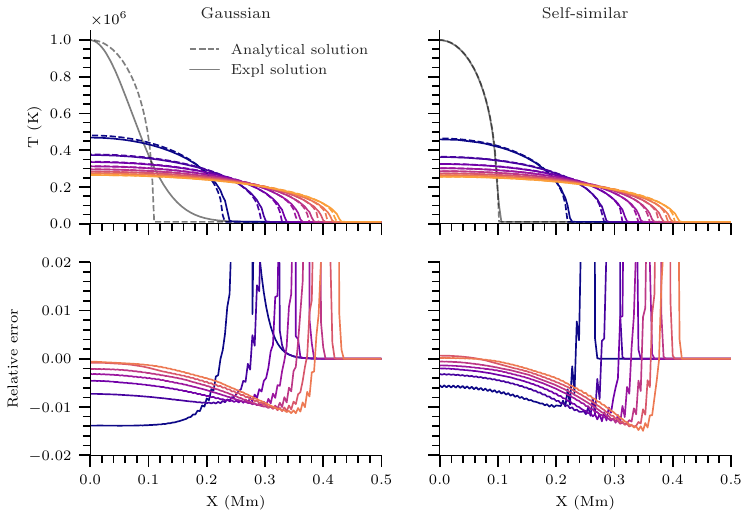}
    \caption{Evolution of the explicit solution for $\Delta t= 10^{-5}\text{ s} $ (solid), compared to the first-order analytical approximation (dashed). Here, the solution for every 0.4 seconds until 3.2 seconds is shown with lines from dark to light. The relative error of the Gaussian profile quickly becomes comparable to that of the self-similar profile for solutions in the $\chi t \gg 1$ regime.}
    \label{fig: ss_comp}
\end{figure*}

From now on, we shall consider primarily the self-similar test. Figure \ref{fig: ss_ic} evidently shows that a Gaussian profile produces acceptable results to compare to a self-similar solution in the $\chi t \gg 1$ regime. In the direction of the magnetic field, the tails steepen to form shock-fronts, as in the self-similar case. The other direction remains completely unchanged in its profile. Therefore, the Gaussian shape provides a better-behaving alternative to the self-similar initial condition, since the profile is naturally continuous everywhere, and does not have as large initial temperature gradients, so less ringing occurs within the solution. Figure \ref{fig: ss_comp} shows that the relative error,
\begin{equation}
    \text{Relative error} = \frac{T_{\text{numerical}} - T_{\text{analytical}}},{T_{\text{analytical}}}
\end{equation}
between the two initial conditions is sufficiently similar for longer timescales. The largest differences between the numerical method and the analytical solutions are at the peak and in the tails. The majority of the error at the tails is a consequence of neglecting higher orders in the approximation \citep{Furuseth2024}, so the best area for analysis and validation of the numerical methods comes from the temperatures near the peak of the profile. It is clear to see the larger instabilities within the self-similar initial condition's evolution, and this leads to a slight decrease in the converged peak temperature, where the Gaussian converges towards a relative error of zero at the peak.\\

\subsection{Time-stepping}
\begin{figure}[h!]
    \centering
    \resizebox{\hsize}{!}{\includegraphics[scale=1.2]{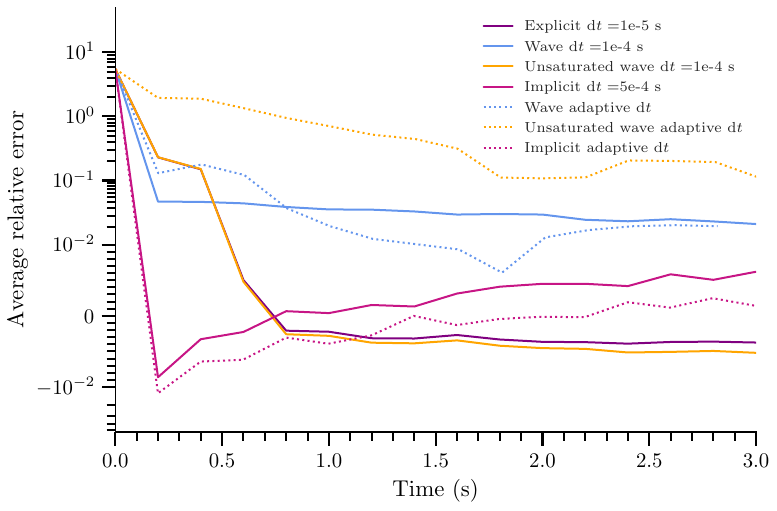}}
    \caption{Average relative error around the peak ($T > 10^5 \text{ K}$) for various set-ups.}
    \label{fig: ss_relerr}   
\end{figure}

\begin{figure}[h!]
    \centering
    \resizebox{\hsize}{!}{\includegraphics[scale=1.2]{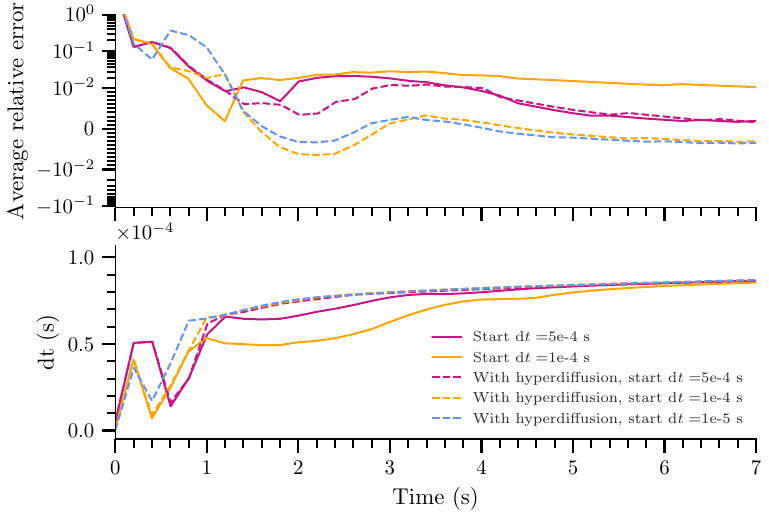}}
    \caption{Average relative error around the peak ($T > 10^5 \text{ K}$) of variable time-stepping in the wave method, using hyper-diffusive terms (dashed).}
    \label{fig: ss_hypdiff}
\end{figure}

Figure \ref{fig: ss_comp} is the result of a stable explicit run, with constant time-step, $\Delta t= 10^{-5}\text{ seconds}$. This is close to the upper limit for a stable explicit $\Delta t$, due to the condition derived in equation \ref{eqn: expl_lim}. On the other hand, the wave method stays stable for time-steps over two orders of magnitude larger, becoming unstable in the self-similar test for constant $\Delta t \approx 3 \times 10^{-3} \text{ seconds}$. As $\Delta t$ increases, the method becomes increasingly wavy\footnote{It is important to note that this waviness is not instability, just a characteristic of the solutions from the nature of the wave method. However, the waves may grow into instabilities if not controlled.} in the initial iterations, and so the magnitude of the relative error is much larger at early times than for the explicit method.
Figure \ref{fig: ss_relerr} around time $t=0.5 \text{ s}$ is an example of this, where the wave method with adaptive time-stepping reaches larger $\Delta t$ than the constant time-step run, and as such, the relative error stays larger for longer. This is also partly due to the larger time-stepping, relating to a larger value of $\tau$, and so a slower rate of convergence. In Fig. \ref{fig: ss_hypdiff}, we see that the adaptive time-step runs all have a period of time where the time-step decreases in order to manage the instability from initial waviness. These smaller $\Delta t$ also allowed the adaptive run in Fig. \ref{fig: ss_relerr} to converge slightly faster than the constant run. The erratic characteristics of the relative error later on in the same run (such as around $t\approx 1.7 \text{ s}$) are due to additional waviness as the $\Delta t$ increases to near the stable limit again. \\

\begin{figure*}
    \centering
    \includegraphics[scale=1.2]{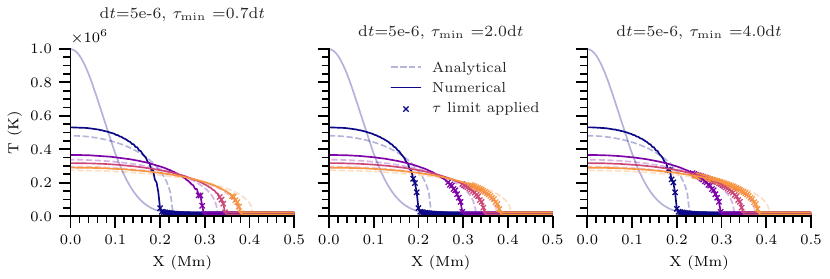}
    \caption{Areas of application to the limit on $\tau$ within the wave method for various limiting values. Solutions are shown for every 0.8 seconds.}
    \label{fig: tau}
\end{figure*}

As with the 1D convergence test, the wave solution lags far behind the analytical solution initially (see Fig. \ref{fig: tau}), and a considerable amount of time passes before the method reaches  errors of the same order as the explicit method; roughly 6 seconds, as is shown in Fig. \ref{fig: ss_hypdiff}. Due to such short time-steps needed, the explicit method would need to perform around 70 000 iterations to reach the point where the saturated wave method becomes similarly accurate, which is computationally expensive and time-consuming. Excluding saturation in the wave method, as is seen in Fig.~\ref{fig: ss_relerr}, performs almost identically regarding the relative error to the explicit method for constant time-stepping. The same cannot be said for the adaptive time-stepping version, which performs the worst of all the methods, due to the stricter conditions for stability needed, as was discussed before. Therefore, although the unsaturated method offers a time-step of an order larger than the explicit method, the requirement for constant time-stepping is still a large time-constraint, and most likely not compatible within larger simulations that need adaptive time-stepping to ease computational cost.\\

\begin{figure}
    \centering
    \resizebox{\hsize}{!}{\includegraphics[scale=1.2]{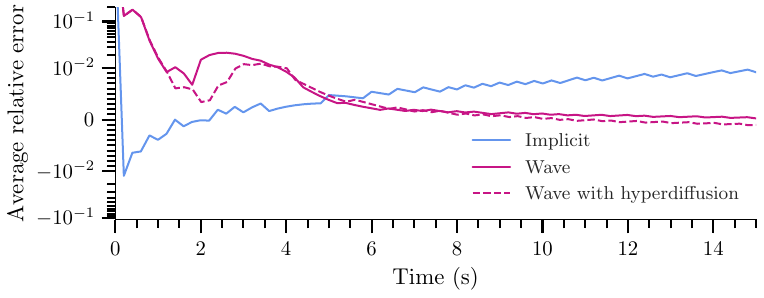}}
    \caption{Long-term error convergence for the wave and implicit methods, with $\Delta t = 5 \times 10^{-4} \text{ seconds}$.}
    \label{fig: impl_err}
\end{figure}

Another alternative would be to use the implicit method, since the implicit error in Fig. \ref{fig: ss_relerr} starts in a similar fashion to the explicit case by quickly decreasing in magnitude from the initial overestimate (on average). However, the solution then overshoots and increases again, this time as an underestimate, before slowly decreasing with time to become an overestimate again. In the long term, the implicit error continues to grow as an overestimate, rather than converging on the analytical solution (see Fig. \ref{fig: impl_err}). This could highlight an issue within the multi-grid, such as over-smoothing or improper weighting between the before- and after-time-step subroutines. It could also just be the build-up in convergence errors if not enough iterations have been used within the multi-grid or relaxation regimes. However, in the period of time before the wave solution has converged, the implicit method provides a robust and stable alternative method which can run at larger $\Delta t$. There is some sacrifice in run time, due to the much larger computational time needed for implicit calculations, but since the maximum $\Delta t$ is over an order greater than for the explicit method, the speed-up is still considerable. \\

The outcome of the variable time-stepping set-up depends significantly on the starting $\Delta t$ value. As is discussed in Sect. \ref{sec: Implem}, the value of $\tau$ is directly proportional to $\Delta t^2$. Choosing a smaller time-step to start will therefore allow the method to converge on the parabolic solution faster, but also expose the solution to more instability. The time-step must increase to more stable $\Delta t$ before it can control this instability, and this takes more iterations for smaller starting $\Delta t$, meaning the method has unstable tendencies for a larger number of oscillations which can grow and crash the run. \verb|Bifrost| contains a series of hyper-diffusion parameters \citep[see][]{Gudiksen2011} to combat such situations. With appropriate hyper-diffusion, the initial instability is sufficiently damped and lower starting time-steps, such as $\Delta t=10^{-5} \text{ s}$ in Fig. \ref{fig: ss_hypdiff} can be achieved. Then, the method converges quickly towards zero relative error due to lower $\tau$ values initially. Notably, the hyper-diffusion completely quashes the secondary error peak seen between 1.4~and~3~seconds, allowing the adaptive $\Delta t$ to continue to increase to the maximum value over  4~simulation~seconds faster than without hyper-diffusion (see Fig. \ref{fig: ss_hypdiff}, bottom panel). We can see in the top panel of Fig. \ref{fig: ss_hypdiff}, that the hyper-diffusive runs starting at $\Delta t = 10^{-4}$ and $10^{-5} \text{ s}$ both converge very well within $7 \text{ seconds}$, with the lower time-step proving the most accurate throughout, whereas the larger starting time-step of $5 \times 10^{-4} \text{ s}$ converges slower due to the overall smaller convergence rate, and larger instabilities.

\begin{figure*}[h]
    \centering
\includegraphics[scale=1.2]{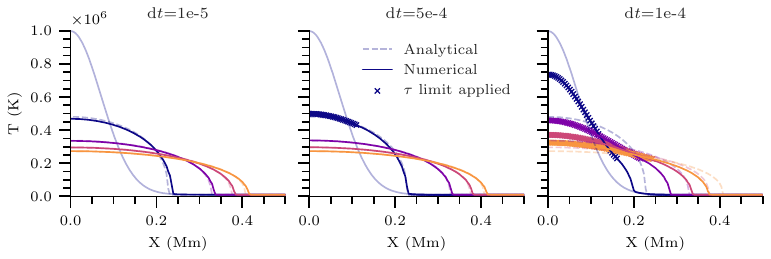}
    \caption{Application of a limit on the heat flux in the explicit method. Solutions are shown every 0.8 seconds.}
    \label{fig: ucmax}
\end{figure*}

\subsection{Safety limits}
In equation \ref{Eqn: tau}, we can see the various terms that can affect the size of the wave convergence factor, $1/\tau$. It should be noted that $\tau$ scales with $T^{5/2}$. This means at the peak of the profile, where $T$ is large, the convergence will be at its slowest, whilst in the tails where $T$ is smaller, the convergence rate can become very high $(\to \infty)$. Here is where we see the activation of the throttle on $\tau$, described in equation \ref{eqn: throttle}. In Fig. \ref{fig: tau}, we test the activation of this limit for $\alpha = (0.7,2.0,4.0)$. As expected, the larger the limit, the more of the temperature profile has this limit applied to it. However, the relative error remains similar, independent of how much of the profile the limit is applied to. Below $\alpha = 0.7$, both the 1D and self-similar simulations crash, due to instability travelling down the steep side of the profiles towards $T_\infty$. In the $\alpha = 0.7$ case, the instability exists in all iterations, but never becomes unmanageable. In the $\alpha=2.0$ case, however, the initial instability is dampened out by time $t=2.4 \text{ s}$, guaranteeing stability from there on. Since there seems to be no distinct loss in accuracy by using a larger $\alpha$, \verb|Bifrost| by default uses $\alpha = 2.0$. At very small $\Delta t$, $\tau_{\min}$ will be much greater than the calculated $\tau$ for the whole profile, and so the limit is applied everywhere. This will significantly reduce the rate of convergence and thus the short- and mid-term accuracy in those $\Delta t$ regimes. However, for variable time-stepping, the proportion of time, both computationally and physically, spent at such small $\Delta t$ is negligible and the accuracy is likely to be recovered quickly as $\Delta t$ increases. \\

In contrast, what limits the time-stepping for the explicit method is the peak of the profile, where $T$, and hence the heat flux, is largest. As explained in Sect. \ref{sec: Implem} it is important that the heat flux does not become larger than the total amount of energy within in the system at any time. If a similar limit would be applied to $\kappa$ in the explicit method, as it was with $\tau$ in the wave method, an issue arises as the $\Delta t$ increases, since then $\kappa_{\max}$,
\begin{equation}
    \kappa_{\max} = \left(0.7\frac{\Delta x^2}{\Delta t}\right) e
\end{equation}
from equation \ref{eqn: expl_lim}, becomes smaller and so the limit must be applied to more of the profile. Unlike the limit on $\tau$, Fig. \ref{fig: ucmax} shows that the application of this limit has a significant effect on the accuracy. This is since this limit directly effects the diffusion rate, whereas the limit on $\tau$ merely dampens the rate at which the method converges to the correct solution. Therefore, this must be the dominant factor that controls the time-step such that the flux is maintained appropriately, and this creates a strict limit on $\Delta t$, as is seen in Fig. \ref{fig: vardt_comp}.

\subsection{Performance}

Figure \ref{fig: vardt_comp} compares the adaptive time-stepping for all methods. It is clear to see that the wave method has the advantage over the explicit method, with an average difference of over one order magnitude between the two time-steps at any point in time. The relationship between the implicit and wave methods is a little more complex. We see a dip in $\Delta t$ for the wave method  around $t = 0.4 \text{ s}$, in order to control the initial waviness. The implicit time-step, however, is free to continue increasing, such that fewer iterations might be needed for the implicit method than the wave method for small $t$. Tables \ref{table:1} and \ref{table:2} provide details for simulation runs using different set-ups and methods at times $t=1.6 \text{ s}$ and $t=3.0 \text{ s}$, respectively. These former time relates to the time at which the time-step of the wave method overtakes that of the implicit method, and the latter to final time in Fig. \ref{fig: vardt_comp}. We can see that the drop in $\Delta t$ in the wave method does allow the implicit method to have fewer iterations for a period of time, as is seen in Table \ref{table:1}, but due to the wave method's faster calculation time, the implicit method always requires more time than the wave method, even with fewer iterations. We can see from the constant time-step runs in Table \ref{table:2}, that the complex machinery of the implicit method makes it around three times slower than the calculations of the wave method. For $t > 1.7 \text{ seconds}$, the adaptive time-step wave method also reaches larger time-steps than the implicit method, and so the disparity between computation time will continue to grow for longer runs. \\

The large jump in $\Delta t$ for all methods in the first half a second is the main advantage of having adaptive time-stepping. Table \ref{table:2} shows that running these methods with constant time-stepping needs a lot more iterations to complete the simulation since  the strict conditions at the initial profile would apply throughout. The speed-up for each method is between 30-50 times faster using adaptive time-stepping, with the greatest speed-up found in the implicit method. The wave method clearly dominates in terms of time to completion, and the inclusion of higher hyper-diffusion allows for further acceleration of the time-step, leading to fewer total iterations. It is also the most perceptive method to changes in the initial time-step, due to the dependency on $\Delta t$ within the convergence rate, $\tau$. As expected, starting with a higher initial time-step leads to a faster run overall, but this is where it is important to balance the priority between computation speed and accuracy, since we have discussed before that the higher starting time-steps lead to a longer convergence with respect to the simulation time. For example, we can calculate from the wave test run with initial $\Delta t  = 5 \times 10^{-4} \text{ s}$, and the hyper-diffusive wave test run with initial $\Delta t = 1 \times 10^{-4} \text{ s}$, using Fig. \ref{fig: ss_hypdiff} and Table \ref{table:2}, that for a $14 \%$ increase in computation time, we gain an approximate $87\%$ reduction in the average relative error at $t=3.0$~simulated~seconds. The initial time-step has much less influence on the implicit method, and since the method converges on the correct solution much faster, the choice of $\Delta t$ at the start is less significant, and will most likely depend on external factors in more complex simulations.

\begin{table*}[h]

\centering
\caption{Completion statistics at time  $t= 1.6 \text{ seconds}.$}
\begin{tabular}{ p{1.5cm} c c c c  }
 \hline
 \hline
 Method& Starting $\Delta t \text{ (s)}$ & Number of iterations & Computation time (s)  & Average time per iteration (s)\\
 \hline
 Explicit &$10^{-5}$ & 6605 & 21.588 & $3.2684 \times 10^{-3}$\\
 Wave* & $10^{-5}$    & 510 & 2.7779&$5.4469 \times 10^{-3}$\\
 Wave    & $10^{-4}$& 522 & 2.8059&$5.3753 \times 10^{-3}$\\
 Wave & $5 \times 10^{-4}$ & 411 & 2.1555&$5.2445 \times 10^{-3}$\\
 Implicit & $ 10^{-4}$ & 391 & 5.2934 &$1.3538 \times 10^{-2}$\\
 Implicit& $5 \times 10^{-4}$  & 387  & 5.2180&$1.3483 \times 10^{-2}$\\
 \hline
\end{tabular}
\tablefoot{Completion statistics for the numerical methods at $t= 1.6 \text{ seconds}$ with adaptive time-stepping, run on a 2D grid with $[256 \times 256]$ grid points. Starred methods have a higher hyper-diffusivity applied. All times are taken for a simulation run on 16 cores in parallel, and exclude processing of output data.}
\label{table:1}
\end{table*}

\begin{table*}[h]
\centering
 \caption{Completion statistics at time $t= 3.0 \text{ seconds}$.}
\begin{tabular}{ p{1.2cm} c c c c c }
 \\
\hline
 \hline
 Method& Starting $ \Delta t \text{ (s)}$ & Constant time-step & Number of iterations & Computation time (s) & Average time per iteration (s)\\
 \hline
 Explicit   & $10^{-5}$ & Y    & 300000 & 968.58& $3.2286 \times 10^{-3}$\\
 Implicit& $10^{-4}$ & Y   & 30001   & 328.96 &  $1.0964 \times 10^{-2} $  \\
 Wave & $10^{-4}$ & Y & 30001    & 111.67& $3.7222\times 10^{-3}$ \\
 \hline
 Explicit & $10^{-5}$&N  & 9057   & 29.506& $3.2578\times 10^{-3}$\\
 Wave* &   $10^{-5}$&N  & 692 & 3.2766& $4.7349\times 10^{-3}$\\
 Wave* & $10^{-4}$&N &  659 &3.2243 & $4.8927 \times 10^{-3}$\\ 
 Wave & $10^{-4}$&N &  774 &3.6749 & $4.7479\times 10^{-3}$\\
 Wave    & $5 \times 10^{-4}$& N& 613 & 2.8094& $4.5830\times 10^{-3}$\\
 Implicit& $10^{-4}$ &N & 577 & 7.3755 & $1.2782 \times 10^{-2}$\\
 Implicit & $5 \times 10^{-4}$ &N & 573 & 7.1232 & $1.2431 \times 10^{-2}$  \\
 \hline
\end{tabular}
\tablefoot{Completion statistics for the numerical methods at $t=3.0 \text{ seconds}$, run on a 2D grid with $[256 \times 256]$ grid points. Starred methods have a higher resistivity applied. All times are taken for a simulation run on 16 cores in parallel, and exclude processing of output data.}
\label{table:2}

\end{table*}

\section{Discussion}\label{sec:concl}

The analysis above has been carried out on pure thermal tests. In other words, all velocities, densities, and magnetic field remain constant throughout. However, the Spitzer equation is, in reality, an addition to the energy equation, which has knock-on effects on the entire magnetohydrodynamics system. It is important to understand how a change in heat flux, such as the wave method's initial waviness, affects an evolving simulation. Similarly, it is important to understand how the MHD quantities interact with and influence the heat flux term. Testing this is outside the scope of this paper, but we can speculate that the inclusion of more physics will most likely create natural stabilisers for conduction, such that the time-stepping range increases and conditions for stability become dependent on factors outside of the \verb|Spitzer| module. For example, with the inclusion of a non-stationary fluid (i.e. that $\mathbf{v} \neq \mathbf{0}$), the initial peak will diffuse faster, relaxing the conditions for stability quicker. Within the wave method, $\tau$ will be greater where the velocities are larger, which ensures that the wave method can still converge appropriately where the characteristics of the plasma may be changing quicker. However, additional physics also introduces more opportunities for shocks, and other extreme phenomena which may negatively impact the thermal conduction code's performance. This would need to be tested in more complex simulations.\\

We have discussed the various methods of calculating thermal conductivity within the parallel MHD solar atmosphere code, \verb|Bifrost|. Any simulation may be run using three different numerical methods for solving the heat flux term. The explicit method, as is expected, provides an accurate but computationally slow solution, since the conditions for stability require very small time-steps. This method is likely only to be used for short simulation runs where the precision of the heat flux term is an important characteristic. As such, it is unlikely to be used for long, realistic solar simulations, as the time-step limit is too taxing. \\

The most beneficial alternative is the saturated wave method from \citet{Rempel2017}. The maximum $\Delta t$ available (allowing stability) is over an order of magnitude larger than for the explicit method, and there is a very low relative error after an initial convergence period. There are two disadvantages to the saturated wave method. Firstly, that the code can become unstable for small $\Delta t$. It is often useful to start or restart a simulation with a very small time-step to let changes in set-up, such as an increased magnetic field, settle to an equilibrium quickly. We have shown that with the help of hyper-diffusion, this instability can be maintained and no accuracy seems to be lost in the long run. The second disadvantage is the convergence time. There is a significant period of time in which the accuracy of the wave method is up to five times worse than the implicit and explicit methods. In realistic simulations, it is important to be aware of this convergence time when analysing the heat flux in the early iterations. This is usually not a problem since most simulations will be run for a set period of time before analysis starts, in order that the simulation be completely self-driven and any artificial phenomena from the initial conditions, of which the waviness of this method could be attributed to, has reduced to minimum levels. We show that a limit on the convergence speed $\tau$ must be implemented for stability, but that has no effect on the accuracy of the method, and negligible effect on the long-term convergence time. We suggest a limit of $\tau = 2.0$, slightly higher than the minimum possible value of $\tau$, $\tau_{\min}=0.7$, in order to avoid unnecessary ringing, but found no reason to increase it further to the extent of \citet{Rempel2017} and \citet{Navarro2022}, who both use $4$. Our limit is still higher than that of \citet{warnecke2020}, where $\tau$ is intrinsically related to the Alfvén speed. Nevertheless, our implementation of the wave method proves to be suitably dynamic and robust, and outperforms both alternative methods in the long term. It is important to consider the sensitivity of this method to the initial time-step, as considerable changes in accuracy may be possible with small alterations to this choice. \\

The implicit method provides an important solution for those needing more accurate medium-range heat flux values. This is the best option for analysis in the region where the wave method is not near convergence, and the explicit method would take too long with short time-steps to calculate. The larger machinery of the implicit method still has a time cost, but this is mostly counteracted by the larger time-steps than are possible for the explicit method, leaving the difference between computation times in the wave and implicit methods as manageable for low numbers of iterations. The simplicity of these tests allowed the multi-grid levels to be set with only a small number of repetitions each and still converge. For more complex simulations, it is possible that the repetition count would need to increase, which would in turn increase the computation time and cause the disparity between the wave and implicit methods to grow. With this in mind, the wave method becomes the most desirable choice for most simulations, with opportunity for the development of a new method that could be used alongside it for analysis in the convergence period.

\begin{acknowledgements}
    
This project has received funding from the European Union's Horizon 2020 research and innovation programme under the Marie Skłodowska-Curie [grant agreement Nº 945371], and is also supported by the Research Council of Norway through its Centres of Excellence scheme, project number 262622.

\end{acknowledgements}

\bibliographystyle{aa}
\bibliography{Main}

\end{document}